\documentstyle[epsf]{elsart}

\def\lQ{\Lambda_{\rm QCD}}
\newcommand{\nn}{\nonumber}
\newcommand{\be}{\begin{equation}}
\newcommand{\ee}{\end{equation}}
\newcommand{\bea}{\begin{eqnarray}}
\newcommand{\eea}{\end{eqnarray}}
\def\al{\alpha}
\def\als{\alpha_{\rm s}}
\def\siml{{\ \lower-1.2pt\vbox{\hbox{\rlap{$<$}\lower6pt\vbox{\hbox{$\sim$}}}}\ }}

\begin{document}
\begin{frontmatter}
\begin{flushright}
\tt{TTP01-27}
\end{flushright}
\vskip 1truecm
\title{Next-to-leading-log renormalization-group running in
heavy-quarkonium creation and annihilation}
\author {Antonio Pineda}
\address{Institut f\"ur Theoretische Teilchenphysik,  
        Universit\"at Karlsruhe, D-76128 Karlsruhe, Germany}

\begin{abstract}
In the framework of potential NRQCD, we obtain the next-to-leading-log
 renormalization-group running of the matching coefficients for the heavy
 quarkonium production currents near threshold. This allows to obtain
 S-wave heavy-quarkonium production/annihilation observables with
 next-to-leading-log accuracy within perturbative QCD. In particular, we
 give expressions for the decays of heavy quarkonium to $e^+e^-$ and to
 two photons. We also compute the $O(m\alpha^8\ln^3\alpha)$ corrections to the
 Hydrogen spectrum.
\end{abstract}

\vspace{1cm}
{\small PACS numbers: 12.38.Cy, 12.38.Bx, 11.10.St, 12.39.Hg}
\end{frontmatter}

\newpage        

\pagenumbering{arabic}
Heavy quark-antiquark systems near threshold are characterized by the
small relative velocity $v$ of the heavy quarks in their center of
mass frame. This small parameter produces a hierarchy of widely
separated scales: $m$ (hard), $mv$ (soft), $mv^2$ (ultrasoft), ... .
The factorization between them is efficiently achieved by using
effective field theories, where one can organize the calculation as
various perturbative expansions on the ratio of the different scales
effectively producing an expansion in $v$. The terms in these series
get multiplied by parametrically large logs: $\ln v$, which can also
be understood as the ratio of the different scales appearing in the
physical system. Again, effective field theories are very efficient in
the resummation of these large logs once a renormalization group (RG)
analysis of them has been performed. This will be the aim of this
paper for annihilation and production processes near threshold.

We will restrict ourselves, in this paper, to the situation where $\lQ
\ll m\als^2$ (to be implicit in what follows), which is likely to be
relevant, at least, for $t$-$\bar t$ production near threshold.

NRQCD \cite{NRQCD} has an ultraviolet (UV) cut-off $\nu_{NR}=\{\nu_p,\nu_s\}$
satisfying $mv \ll \nu_{NR} \ll m$. At this stage $\nu_p \sim
\nu_s$. $\nu_p$ is the UV cut-off of the relative three-momentum of
the heavy quark and antiquark, ${\bf p}$. $\nu_s$ is the UV cut-off of the
three-momentum of the gluons and light quarks.

Potential NRQCD (pNRQCD) \cite{pNRQCD} (see \cite{pNRQED,long} for
details) is defined by its particle content and cut-off
$\nu_{pNR}=\{\nu_p,\nu_{us}\}$, where $\nu_p$ is the cut-off of the
relative three-momentum of the heavy quarks and $\nu_{us}$ is the cut-off of the
three-momentum of the gluons and light quarks. They satisfy the following 
inequalities: $|{\bf p}| \ll \nu_p \ll m$ and ${\bf p}^2/m \ll
\nu_{us} \ll |{\bf p}|$. Note that no gluons or light quarks with
momentum of $O(|{\bf p}|)$ are kept dynamical in pNRQCD. The motivation to integrate 
out these degrees of freedom is that they do not appear
as physical (on-shell) states near threshold. Nevertheless, they can appear off-shell 
and, since their momentum is of the order of the relative three-momentum of the heavy 
quarks, integrating them out produces non-local terms (potentials) in three-momentum
space. Indeed, these potentials encode the non-analytical behavior in the transfer
momentum of the heavy quark, ${\bf k}={\bf p}-{\bf p}'$, of the order
of the relative three-momentum of the heavy quarks.

The matching process, which basically means the computation of the
potentials, is carried out for a given external incoming (outcoming)
momentum ${\bf p}$ (${\bf p}'$). Therefore, one has to sum over all of them in
the pNRQCD Lagrangian, since they are still physical degrees of
freedom as far as their momentum is below $\nu_p$. In position space, 
this means that an integral over ${\bf x}$, the relative distance
between the heavy quarks, appears in the Lagrangian when written in
terms of the heavy quark-antiquark bilinear.

Within pNRQCD, integrals over ${\bf p}$ (or ${\bf x}$) appear when
solving the Schr\"odinger equation that dictates the dynamics of the
heavy quarkonium near threshold. At lower orders, these integrals are
finite effectively replacing ${\bf p}$ by $\sim m\als$. Nevertheless,
at higher orders in quantum mechanics perturbation theory and/or if
some singular enough operators are introduced (as it will be the case
of the heavy quarkonium production currents) singularities
proportional to $\ln\nu_p$ appear. These must be absorbed by the potentials
or by the matching coefficients of the currents. We will describe how
to resum the logarithms associated to this cutoff within pNRQCD.

A RG analysis for non-relativistic systems have been addressed before
in Refs.  \cite{LMR,vNRQCD2,V1MS}, where they match to an effective
theory called vNRQCD. On physical terms, this theory should be
equivalent to the previously defined pNRQCD once the RG evolution has
been performed and the soft degrees of freedom have been integrated out,
as only ultrasoft gluons and light fermions and potential quarks are
left as dynamical degrees of freedom. We will compare with their
results. In some cases disagreement will be found.

Let us now describe the matching between QCD and pNRQCD within an RG
framework. For the case where no divergences proportional to $\ln\nu_p$
appear, the procedure reduces to the results of Ref. \cite{RGmass} to
which we refer for the notation and background material  necessary
to follow this paper.

We first address the procedure that gives the running of the potentials.
One first does the matching from QCD to NRQCD. The latter depends on
some matching coefficients: $c(\nu_s)$ and $d(\nu_p,\nu_s)$, which can
be obtained order by order in $\als$ (with $\nu_p=\nu_s$) following the
procedure described in Ref. \cite{Manohar}. The $c(\nu_s)$ stand for the
coefficients of the operators that already exist in the theory with only
one heavy quark (ie. HQET) and the $d(\nu_p,\nu_s)$ stand for the
coefficients of the four heavy fermion operators. The starting point of
the renormalization group equation can be obtained from these
calculations by setting $\nu_p=\nu_s=m$ (up to a constant of order
one). In principle, we should now compute the running of $\nu_p$ and
$\nu_s$. The running of the $c(\nu_s)$ can be obtained using HQET
techniques \cite{HQET}. The running of the $d(\nu_p,\nu_s)$ is more
complicated. At one-loop, $\nu_p$ does not appear and we effectively
have $d(\nu_p,\nu_s)\simeq d(\nu_s)$, whose running can also be obtained
using HQET-like techniques \cite{RGmass}. At higher orders, the
dependence on $\nu_p$ appears and the running of the $d(\nu_p,\nu_s)$
becomes more complicated. Fortunately, we need not compute the
running of $d$ in this more general case because, as we will see, the
relevant running of the $d$ for near threshold observables can be
obtained within pNRQCD.

The next step is the matching from NRQCD to pNRQCD. The latter depends
on some matching coefficients (potentials). They typically have the
following structure: ${\tilde V}(c(\nu_s), d(\nu_p,\nu_s),\nu_s,\\
\nu_{us},r)$. After matching, any dependence on $\nu_s$ disappears since
the potentials have to be independent of $\nu_s$. Therefore, they could
be formally written as ${\tilde V}(c(1/r), d(\nu_p,1/r),
1/r,\nu_{us},r)$. These potentials can be obtained order by order in
$\als$ following the procedure of Refs. \cite{pNRQCD,pNRQED,long}. The integrals in
the matching calculation would depend on a factorization scale $\mu$,
which should correspond either to $\nu_s$ or to $\nu_{us}$. In the explicit calculation, 
they could be distinguished by knowing the UV and infrared (IR) behavior
of the diagrams: UV divergences are proportional to $\ln\nu_s$, which
should be such as to cancel the $\nu_s$ scale dependence inherited from
the NRQCD matching coefficients, and IR divergences to $\nu_{us}$. In
practice, however, as far as we only want to perform a matching
calculation at some given scale $\mu=\nu_s=\nu_{us}$, it is not
necessary to distinguish between $\nu_s$ and $\nu_{us}$ (or if working
order by order in $\als$ without attempting any log resummation).

Before going into the rigorous procedure to obtain the RG equations of
the potentials, let us first discuss their structure on physical
grounds. As we have mentioned, the potential is independent of
$\nu_s$. The independence of the potential with respect $\nu_s$ allows
us to fix the latter to $1/r$ that, in a way, could be understood as the
matching scale for $\nu_s$\footnote{In practice, the potential is often
first obtained in momentum space so that one could then set
$\nu_s=k$. Note, however, that this is not equivalent to fix $\nu_s=1/r$, since
finite pieces will appear after performing the Fourier
transform.}. Therefore, $1/r$, the point where the multipole expansion
starts, would also provide with the starting point of the
renormalization group evolution of $\nu_{us}$ (up to a constant of order
one).  The running of $\nu_{us}$ can then be obtained following the
procedure described in Refs. \cite{RG,RGmass}.  At the end of the day, we
would have ${\tilde V}(c(1/r),d(\nu_p,1/r),1/r,\nu_{us},r)$, where the
running on $\nu_{us}$ is known and also the running in $1/r$ if the $d$
is $\nu_p$-independent. So far, the only explicit dependence of the
potential on $\nu_p$ appears in the $d$. Nevertheless, the potential is
also implicitly dependent on the three-momentum of the heavy quarks
through the requirement $1/r \sim {\bf p} \ll \nu_p$, and also through
$\nu_{us}$, since $\nu_{us}$ needs to fulfill ${\bf p}^2/m \ll \nu_{us}
\ll |{\bf p}|$. This latter requirement holds if we fix
$\nu_{us}=\nu_p^2/m$ (this constraint tells you how much you can run
down $\nu_{us}$ in the potential before finding the cutoff $\nu_p^2/m$
caused by the cutoff of ${\bf p}$).

Within pNRQCD, the potentials should be introduced in the
Schr\"odinger equation. This means that integrals over the relative
three-momentum of the heavy quarks take place. When these integrals
are finite one has ${\bf p} \sim 1/r \sim m\als$ and ${\bf p}^2/m \sim
m\als^2$. Therefore, one can lower $\nu_{us}$ down to $\sim m\als^2$
reproducing the results obtained in Ref. \cite{RGmass}. In some cases,
in particular in heavy quarkonium creation, the integrals over ${\bf
p}$ are divergent, and the log structure is dictated by the ultraviolet
behavior of ${\bf p}$ and $1/r$. This means that we can not replace
$1/r$ and $\nu_{us}$ by their physical expectation values but rather by
their cutoffs within the integral over ${\bf p}$. Therefore, for the
RG equation of $\nu_p$, the anomalous dimensions will depend (at
leading order) on ${\tilde
V}(c(\nu_p),d(\nu_p,\nu_p),\nu_p,\nu_p^2/m,\nu_p)$\footnote{Roughly speaking, this result 
can be
thought as expanding $\ln r$ around $\ln \nu_p$ in the potential ie.
\bea
\nn 
{\tilde V}(c(1/r),d(\nu_p,1/r),1/r,\nu_p^2/m,r) &\simeq& {\tilde
V}(c(\nu_p),d(\nu_p,\nu_p),\nu_p,\nu_p^2/m,\nu_p)
\\
&&
+\ln(\nu_pr)r{d \over d r} {\tilde V}\bigg|_{1/r=\nu_p} + \cdots 
\,.
\eea 
The $\ln(\nu_pr)$ terms may give subleading contributions to the
anomalous dimension when introduced in divergent integrals over ${\bf
p}$. The discussion at this stage is not very rigorous and a more
precise discussion would require a full detailed study within
dimensional regularization, which goes beyond the aim of this work. 
Nevertheless, we do not expect it to
change the underlying idea, although it deserves
further investigations.} and the running will go from $\nu_p \sim m$
down to $\nu_p \sim m\als$. Note that, at this stage, a single cutoff,
$\nu_p$, exists and the correlation of cutoffs can be seen. The
importance of the idea that the cutoffs of the non-relativistic
effective theory should be correlated was first realized by Luke,
Manohar and Rothstein in Ref. \cite{LMR} (for an application to QED see \cite{QED}). Note 
also that at the matching scale $\nu_p \sim m$, 
what it would be the ultrasoft cutoff is also of order $m$. In this
sense it should be understood the statement in Ref. \cite{LMR} that
ultrasoft gluons appear at the scale $m$, a point that becomes relevant
within a RG approach.

With the above discussion in mind, the matching between NRQCD and
pNRQCD could be thought as follows. One does the matching by computing
the potentials order by order in $\als$ at the matching scale
$\nu_p=\nu_s=\nu_{us}$ following the procedure of Refs. \cite{pNRQCD,pNRQED,long} (by
doing the matching at a generic $\nu_p$ some of the running is
trivially obtained).  The structure of the potential at this stage
then reads ${\tilde V}(c(\nu_p),d(\nu_p,\nu_p),\nu_p,\nu_p,\nu_p)$ (and
similarly for the derivatives with respect $\ln r$ of the
potential). This provides the starting point of the renormalization
group evolution of $\nu_{us}$ (up to a constant of order one).  The
running of $\nu_{us}$ can then be obtained following the procedure
described in Refs. \cite{RG,RGmass}. For the final point of the evolution
of $\nu_{us}$, we choose $\nu_{us}=\nu_p^2/m$. At the end of the day, we
obtain ${\tilde V}(c(\nu_p), d(\nu_p,\nu_p),\nu_p,\nu_p^2/m,\nu_p)
\equiv {\tilde V}(\nu_p)$.

The running of $\nu_p$ goes from $\nu_p=m$ (this was fixed when the
matching between QCD and NRQCD was done) up to the physical scale of
the problem $\nu_p \sim m\als$. If the running of the NRQCD
matching coefficients is known, the above result gives the complete running of
the potentials. The procedure to get the running of the $c$ is known
at any finite order. For the $d$ it is just known at one-loop order, 
since, at this order, it is only $\nu_s$-dependent. Nevertheless, at
higher orders, dependence on $\nu_p$ appears. Therefore, the above
method is not complete unless an equation for the running of $\nu_p$ is
provided. This is naturally given within pNRQCD. It appears through the
iteration of potentials. Let us consider this situation more in
detail. We first remind what the Hamiltonian in pNRQCD for the
singlet sector is (see Ref. \cite{RGmass} for notation and further details):
\bea
h_s&=&c_k{{\bf p}^2 \over m}-c_4{{\bf p}^4 \over 4m^3}- C_f {\alpha_{V_{s}} \over r}-{C_fC_A
D^{(1)}_s \over 2mr^2}
\nn
\\
&&- { C_f D^{(2)}_{1,s} \over 2 m^2} \left\{ {1 \over r},{\bf p}^2 \right\}
+ { C_f D^{(2)}_{2,s} \over 2 m^2}{1 \over r^3}{\bf L}^2
+ {\pi C_f D^{(2)}_{d,s} \over m^2}\delta^{(3)}({\bf r})
\nn
\\
& & + {4\pi C_f D^{(2)}_{S^2,s} \over 3m^2}{\bf S}^2 \delta^{(3)}({\bf r})
+ { 3 C_f D^{(2)}_{LS,s} \over 2 m^2}{1 \over r^3}{\bf L} \cdot {\bf S}
+ { C_f D^{(2)}_{S_{12},s} \over 4 m^2}{1 \over r^3}S_{12}(\hat{\bf r})
\,,
\eea
where $C_f=(N_c^2-1)/(2N_c)$ and we will set $c_k=c_4=1$ (we will only eventually use $c_4$ for
tracking of the contribution due to this term). The propagator of the
singlet is (formally)
\be
{1 \over \displaystyle{E-h_s}}
\,.
\ee
At leading order (within an strict expansion in $\als$) the propagator of the singlet reads
%%%%%%%%%%%%%%%%%%%%%%%%%%%%%%%%%%%%%%%%%%%%%%%%%%%%%%%%%%%%%%%%%%%%%%%%%%%%%%%%%%
\begin{figure}[htb]
\makebox[0.5cm]{\phantom b}
\epsfxsize=3truecm \epsfbox{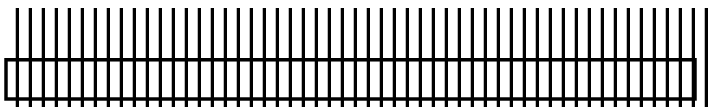}
\put(25,1){$
\label{Gc}
=G_c(E)=\displaystyle{{1 \over \displaystyle{E-h_s^{(0)}}}
=
{1 \over \displaystyle{E-{\bf p}^2/m-C_f \als/ r}}}
\,.$}
\end{figure}
%%%%%%%%%%%%%%%%%%%%%%%%%%%%%%%%%%%%%%%%%%%%%%%%%%%%%%%%%%%%%%%%%%%%%%%%%%%%%%%%%%

If we were interested in computing the spectrum at $O(m\als^6)$, one
should consider the iteration of subleading potentials ($\delta h_s$) in the
propagator as follows:
\be
G_c(E)\delta h_s G_c(E) \cdots \delta h_s G_c (E)
\,.
\ee
In general, if these potentials are singular enough, these
contributions will produce logarithmic divergences due to potential
loops. These divergences can be absorbed in the
matching coefficients, $D^{(2)}_{d,s}$ and $D^{(2)}_{S^2,s}$, of the 
local potentials (those proportional to the $\delta^{(3)}({\bf r}$)) providing
with the renormalization group equations of these matching
coefficients in terms of $\nu_p$. Let us explain how it works in
detail. Since the singular behavior of
the potential loops appears for $|{\bf p}| \gg \als/r$, a perturbative
expansion in $\als$ is licit in $G_c(E)$, which can be approximated by

%%%%%%%%%%%%%%%%%%%%%%%%%%%%%%%%%%%%%%%%%%%%%%%%%%%%%%%%%%%%%%%%%%%%%%%%%%%%%%%%%%
\begin{figure}[htb]
\makebox[0.5cm]{\phantom b}
\epsfxsize=3truecm \epsfbox{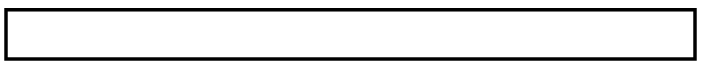}
\put(25,1){$
\label{Gc0}
=G_c^{(0)}(E)=\displaystyle{
{1 \over \displaystyle{E-{\bf p}^2/m}}}
\,.$}
\end{figure}
%%%%%%%%%%%%%%%%%%%%%%%%%%%%%%%%%%%%%%%%%%%%%%%%%%%%%%%%%%%%%%%%%%%%%%%%%%%%%%%%%%

Therefore, with the accuracy we aim at in this paper, a practical 
simplification follows from the fact that the Coulomb potential, $-C_f
{\als \over r}$, can 
be considered to be a perturbation as far as the computation of the
$\ln \nu_p$ ultraviolet 
divergences is concerned. This means that the computation of the
anomalous dimension can be organized within an expansion in
$\als$ and using the free propagators $G_c^{(0)}$. Moreover, each
$G_c^{(0)}$ produces a potential loop and one extra power of $m$ in the
numerator, which kills the powers in $1/m$ of the different
potentials. This allows the mixing of potentials with different powers
in $1/m$. One typical example would be the diagram in
Fig. 1.\footnote{The diagram in Fig. 1 is also
the relevant one in order to obtain the $O(m\al^8\ln\al^3)$
contribution to the Hydrogen spectrum. See the Appendix.}
 The computation of this diagram would go as follows:
\be
{1 \over \displaystyle{E-{\bf p}^2/m}}
{\pi C_f D^{(2)}_{d,s} \over m^2}\delta^{(3)}({\bf r})
{1 \over \displaystyle{E-{\bf p}^2/m}}
C_f {\al_{V_s} \over r}
{1 \over \displaystyle{E-{\bf p}^2/m}}
{\pi C_f D^{(2)}_{d,s} \over m^2}\delta^{(3)}({\bf r})
{1 \over \displaystyle{E-{\bf p}^2/m}}
\,.
\ee
Using $\delta^{(3)}({\bf r})=|{\bf r}=0\rangle\langle{\bf r}=0|$, we can see that the 
relevant computation reads (instead of $\al_{V_s}$ one could use
$\als$ since the non-trivial running of $\al_{V_s}$ is a subleading
effect. Nevertheless, we keep $\al_{V_s}$ since it allows to keep track of the
contributions due to the Coulomb potentials)
\bea
&&
\langle{\bf r}=0|
{1 \over \displaystyle{E-{\bf p}^2/m}}
C_f {\al_{V_s} \over r}
{1 \over \displaystyle{E-{\bf p}^2/m}}
|{\bf r}=0\rangle
\nn
\\
&&
\qquad
\sim 
\int \frac{
{\rm d}^d p' }{ (2\pi)^d } \int \frac{ {\rm d}^d p }
{ (2\pi)^d } \frac{ m }{{\bf p}'^2 - mE } 
C_f
\frac{ 4\pi\alpha_{V_s} }{ {\bf q}^2 } \frac{ m }{{\bf p}^2-m E } 
\sim
- C_f\frac{m^2\alpha_{V_s}}{16\pi}  
\frac{ 1 }{\epsilon },
\eea
where $D=4+2\epsilon$ and ${\bf q}={\bf p}-{\bf p}'$. This divergence is absorbed in $D_{d,s}^{(2)}$ contributing to its running at
next-to-leading-log (NLL) order as follows 
\be
\label{eqDd}
\nu_p {d \over d\nu_p}D_{d,s}^{(2)}(\nu_p) \sim 
\alpha_{V_s}(\nu_p)D_{d,s}^{(2)2}(\nu_p)+\cdots
\,.
\ee
Therefore, even without knowing the running of the $d$ 
(which need to be known at NLL order in this case), we
can obtain the running of the potential (one can also think of trading
Eq. (\ref{eqDd}) into an equation for $d$, which is the only unknown 
parameter within the potential). This is so because $D_{d,s}^{(2)}$ is 
only needed with LL accuracy in the right-hand side of Eq. (\ref{eqDd}). 

%%%%Figure Ddnup%%%%
\medskip
\begin{figure}[h]
\hspace{-0.1in}
\epsfxsize=2.8in
\centerline{\epsffile{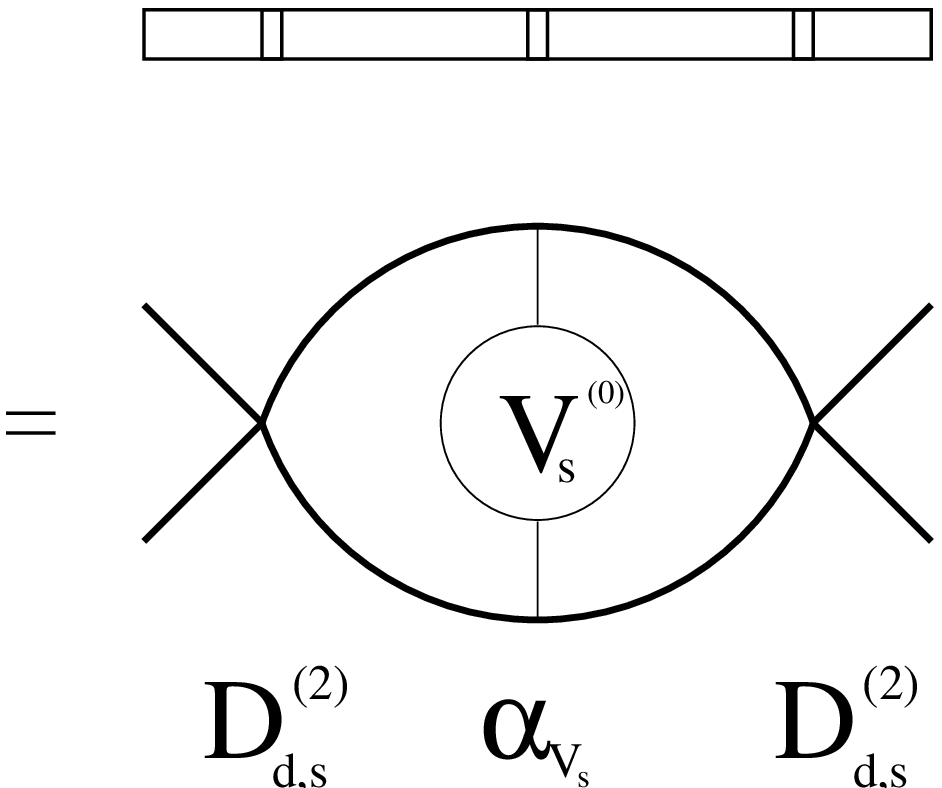}}
\caption {{\it One possible contribution to the running of $D_{d,s}^{(2)}$ at
NLL. The first picture represents the calculation in 
terms of the free quark-antiquark propagator $G_c^{(0)}$ and the small 
rectangles the potentials. The picture
below is the representation within a more standard diagrammatic
interpretation in terms of quarks and antiquarks. The delta potentials
are displayed as local interactions and the Coulomb potential as an
extended in space (but not in time) object.}}
\label{obs12}
\end{figure}
%%%%End Figure Ddnup%%%%

The above method deals with the resummation of logs due to the hard, soft and ultrasoft 
scales. Nevertheless, for some specific kinematical situations even smaller scales could 
appear. Their study, however, goes beyond the aim of this paper. In any case, pNRQCD can 
be considered to be the right starting point to study these kinematical situations. 

The matching scale between QCD and NRQCD is $\nu_p \sim \nu_s \sim
m$. On the other hand, the matching scale between NRQCD and pNRQCD is
also the hard scale: $\nu_p\sim \nu^2_p/m \sim m$. Therefore, one could
wonder about the necessity of using the intermediate theory NRQCD. This
is indeed the attitude in Refs. \cite{LMR,QED,vNRQCD2,V1MS}, where they
directly perform the matching between QCD to an effective field theory:
vNRQCD that, once the RG evolution has been performed and the soft
degrees of freedom have been integrated out, should be physically
equivalent to pNRQCD with $\nu_p \sim m\als$. One motivation for going
through NRQCD is that it allows to perform the factorization of the hard
scale within an effective field theory framework. In fact, a full
factorization of the different regions of momentum that ought to be
integrated out is achieved within pNRQCD. This extremely simplifies the
matching process since one deals with only one scale (region of
momentum) in the loops at each step. In the matching between QCD and
NRQCD only hard loops need to be considered, whereas in the matching
between NRQCD and pNRQCD only soft loops need to be
considered. Moreover, the structure of the UV cutoffs of the theory is
better understood in this way. For instance, one can see that all the
{\it explicit} dependence of the potentials on $\nu_p$ is inherited from
the $d$ matching coefficients. Within a diagrammatic approach the
factorization of the different regions of momentum have been achieved
using the threshold expansion \cite{BS}.

Let us now consider the case of the electromagnetic current, which will
provide an example where to apply the above discussion. 
The
procedure is analogous to the potentials. We first do the matching
from QCD to NRQCD:
\be
{\bar Q}\gamma^{\mu}Q(0) \bigg|_{\rm QCD} \dot= b_{1,\rm 
NR}\psi^\dagger\sigma^i\chi(0)+O(1/m)\bigg|_{\rm NRQCD}\,.
\ee
We will just concentrate in the coefficient $b_{1,\rm NR}$. Within NRQCD, it should
be understood as a function of $\nu_p$ and $\nu_s$,
ie. $b_{1,\rm NR}(\nu_p,\nu_s)$. One should first obtain the matching conditions
at the hard scale. This has been
computed up to two loops \cite{BSSCM} but we will only need the one-loop
expression \cite{Barbieri}:
\be
b_{\rm 1, NR}(m,m)=1-2C_f{\als(m) \over \pi}
,
\ee
since we only aim to a NLL resummation in this paper. If we compare with the previous 
discussion of the potentials, the
matching coefficients $d$ play the role of $b_1$. Therefore, within
pNRQCD, we will need $b_{\rm 1, NR}(\nu_p,\nu_p)\equiv b_{\rm 1, NR}(\nu_p)$. We 
first have to consider $b_{\rm 1, NR}(\nu_p,\nu_s)$. In this case, unlike for the 
$d$'s, there is no running due to $\nu_s$ at the order of interest. This can be easily 
seen in the Coulomb gauge. Moreover, the matching 
from NRQCD to pNRQCD creates the
potentials but let $b_1$ unchanged since soft loops or HQET-like
calculations give zero correction to $b_1$ at the order of interest. Formally, 
\be
b_{\rm 1, NR}\psi^\dagger\sigma^i\chi(0)\bigg|_{\rm NRQCD}=B_{\rm 1, 
pNR}\psi^\dagger\sigma^i\chi(0)\bigg|_{\rm pNRQCD}\,,
\ee
or, in other words, the matching condition reads 
$B_{\rm 1, pNR}(b_{\rm 1, NR}(\nu_p),\nu_{us}=\nu_p)=b_{\rm 1, NR}(\nu_p)$. The running
of $\nu_{us}$ is also trivial as there is none at the order of interest 
(this has to do with the
fact that we are dealing with an annihilation process). Therefore, we
finally have $B_1(\nu_p) \equiv
B_{\rm 1, pNR}(b_{\rm 1, NR}(\nu_p),\nu_p^2/m)=b_{\rm 1, NR}(\nu_p)$.  We can see that we 
are in the analogous
situation to the running of $D_{d,s}^{(2)}(\nu_p)$ versus the running of
$d(\nu_p,\nu_p)$. We now need the RG
equation for $B_1(\nu_p)$. This demands to obtain the
ultraviolet corrections to the current within pNRQCD keeping track of
the contributions due to the different potentials. Fortunately, this
calculation has already been done and we can extract the relevant information from
Ref. \cite{KP2}. The computation goes along the same lines than in the example of 
Fig. 1. The explicit diagrams to be computed for the RG running of
$B_s(\nu_p)$ are given in Fig. 2 (where $s$ denotes the spin). From this
figure, we can clearly illustrate the structure of the
computation. $O(1/m)$ corrections to $h_s^{(0)}$ only need one potential loop to kill
the $1/m$ coefficient. $O(1/m^2)$ corrections to $h_s^{(0)}$  need two
potential loops to kill the $1/m^2$ coefficient and so on. In the
situation with more than one potential loop, the additional potential
loops can be produced without additional $1/m$ factors coming from the
potential only if Coulomb potentials are introduced. This explains why the
$1/m$ potential needs zero Coulomb potential insertions, the $1/m^2$
potentials need one
Coulomb potential insertions and the $1/m^3$ term needs two Coulomb
potential insertions (for the running of $D_{d,s}^{(2)}$ and
$D_{S^2,s}^{(2)}$ we expect a similar structure). In principle, this
would be be a never ending story unless there is an small parameter
that tells us how far we have to go in the calculation in order to achieve
some given accuracy. This is indeed so. The $1/m$ potential is a NLL
effect \cite{RGmass} and therefore higher powers in $D_{s}^{(1)}$ produce NNLL effects
or beyond. On the other hand, the introduction of Coulomb potentials
brings powers in $\als$, which suppresses the order of the calculation. In
our case, for a NLL calculation, the maximum power of the anomalous
dimension should be $\als^2$. This means that with zero $\al_{V_s}$
insertions ($O(1/m)$ potentials) the matching coefficient
($D_{s}^{(1)}$) has to be known with NLL accuracy, with one $\al_{V_s}$ insertion ($O(1/m^2)$ potentials) the
matching coefficients ($D^{(2)}$) have to be
known with LL accuracy and with two $\al_{V_s}$ insertions ($O(1/m^3)$
potentials) the matching coefficients must have no running (this
explains why only $c_4$ is considered at this order). 

%%%%Figure Bs%%%%
\medskip
\begin{figure}[h]
\hspace{-0.1in}
\epsfxsize=2.8in
\centerline{\epsffile{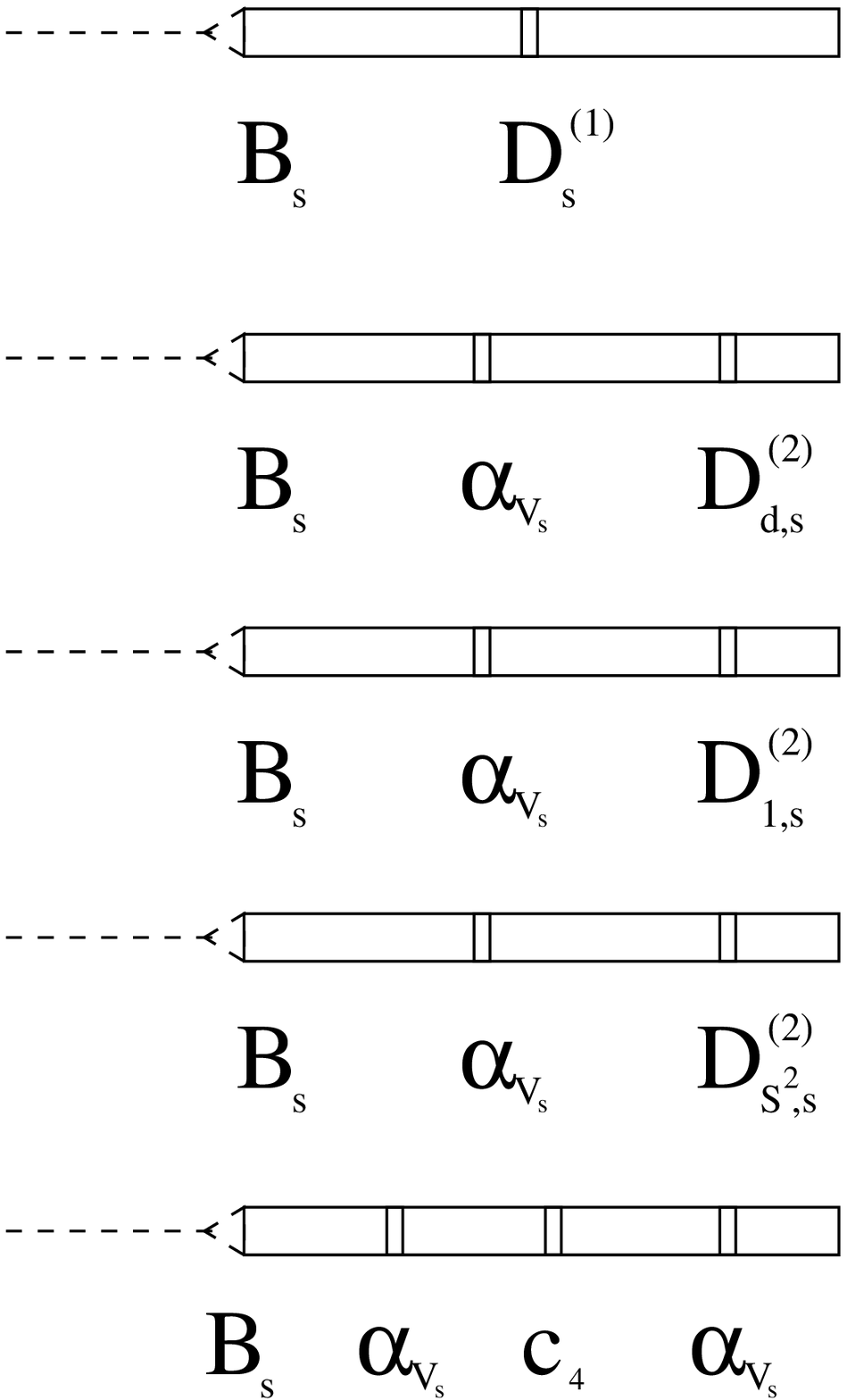}}
\caption {{\it Diagrams, up to permutations, that contribute to the
running of $B_s$.}}
\label{Bs}
\end{figure}
%%%%End Figure Bs%%%%

From the above discussion, the RG equation reads\footnote{The RG equations of $B_s$ within vNRQCD has 
been computed in Ref. \cite{LMR,V1MS}. 
In principle, they are different. Nevertheless, it may happen that fields redefinitions 
of the potentials may make them equal. We have not checked that but it is plausible since, 
as we will see, for the leading and subleading logs (but not beyond) both calculations 
will agree.} 
\be 
\label{cseq}
\nu_p {d  \over d\nu_p} B_s =B_s
\left[
-{C_AC_f \over 2}D_s^{(1)}
-{C_f^2 \over 4}\al_{V_s}
	\left\{\al_{V_s}-{4\over 3}s(s+1)D_{S^2,s}^{(2)}
		-D_{d,s}^{(2)}+4D_{1,s}^{(2)}
	\right\} 
\right]
\,, 
\ee
where $C_A=N_c$, and the RG-improved matching coefficients of the potentials 
can be read from 
Ref. \cite{RG,RGmass} with the assignment $1/r \rightarrow \nu_p$ and $\nu_{us}
\rightarrow \nu^2_p/m$ (see also \cite{pot,long} for calculations of the potentials
at finite orders in $\als$). We have kept the spin explicit so our
results will also be valid for the pseudoscalar current:
\be
{\bar Q}\gamma^{0}\gamma_5Q(0) \bigg|_{\rm QCD}\dot= b_{0,\rm
NR}\psi^\dagger\chi(0)+O(1/m)\bigg|_{\rm NRQCD}\,, 
\ee
with the matching condition \cite{Barbieri2}:
\be
b_0 (m)=1+\left({\pi^2 \over 4}-5\right){C_f \over 2}{\als(m) \over \pi}
.
\ee

Eq. (\ref{cseq}) gives subleading effects
within an strict expansion in $\als$. Therefore, it can be approximated to 
\be 
\label{cseqap}
\nu_p {d  \over d\nu_p} B_s =
-{C_AC_f \over 2}D_s^{(1)}
-{C_f^2 \over 4}\als
	\left\{\als-{4\over 3}s(s+1)D_{S^2,s}^{(2)}
		-D_{d,s}^{(2)}+4D_{1,s}^{(2)}
	\right\} 
\,,
\ee
and the solution reads
\bea 
B_s(\nu_p)&=&b_s(m)+ 
	A_1 {\als(m) \over w^{\beta_0}} \ln(w^{\beta_0})+ 
	A_2 \als(m) \bigg[z^{\beta_0}-1\bigg] + 
	A_3 \als(m) \bigg[z^{\beta_0-2 C_A}-1  \bigg]
\nn
\\
&&+
	A_4 \als(m) \bigg[z^{\beta_0-13C_A/6}-1 \bigg]+
	A_5 \als(m) \ln(z^{\beta_0})
\,,
\label{cssln}
\eea
where $\beta_0={11 \over 3}C_A -{4 \over 3}T_Fn_f$, $z=\left[{\als(\nu_p) \over 
\als(m)}\right]^{1 \over
\beta_0}$ and $w=\left[{\als(\nu_p^2/m) \over \als(\nu_p)}\right]^{1 \over
\beta_0}$. The coefficients $A_i$ in Eq.~(\ref{cssln}) read
\begin{eqnarray} 
\label{Acoeffs}
  A_1 &=& {8\pi C_f \over 3\beta_0^2 }\left(C_A^2 +2C_f^2+3 C_fC_A
  \right) \,,\nn\\ 
  A_2 &=& { \pi C_f [3 \beta_0(26C_A^2+19C_A C_f-32C_f^2)-
     C_A(208 C_A^2+651 C_A C_f+116 C_f^2)]\over 78\,\beta_0^2\, C_A } \,, \nn\\
  A_3 &=& -{\pi C_f^2 \Big[  \beta_0 (4s(s+1)-3)+ C_A (15-14s(s+1)) \Big]
     \over 6 (\beta_0-2 C_A)^2 }\,,\nn\\
  A_4 &=& {24 \pi C_f^2 (3\beta_0-11 C_A)(5 C_A+8 C_f) \over 13\, C_A
     (6\beta_0-13 C_A)^2}\,,\nn\\
  A_5 &=& {-\pi C_f^2 \over \beta_0^2\, (6\beta_0-13C_A) (\beta_0-2C_A)}
      \, \Bigg\{ C_A^2(-9C_A+100C_f)
	\nn\\ 
	&&+\beta_0 C_A(-74C_f+C_A(42-13s(s+1)))
	+6\beta_0^2(2C_f+C_A(-3+s(s+1)))\Bigg\}
     \,.
\end{eqnarray}
Our evaluation can be compared with the result obtained using the
vNRQCD formalism \cite{V1MS}. We agree for the spin-dependent terms
but differ for the spin-independent ones. The disagreement still holds
if we consider QED with light fermions ($C_f \rightarrow 1$, $C_A
\rightarrow 0$, $T_F \rightarrow 1$). Agreement is found if we
consider QED without light fermions ($C_f \rightarrow 1$, $C_A
\rightarrow 0$, $n_f \rightarrow 0$, $T_F \rightarrow 1$). If we
expand our results in $\als$, we can compare with earlier results in
the literature. By following the discussion in Ref. \cite{V1MS}, we
can relate our results with the correction to the wave-function at the
origin as defined in Ref. \cite{KP2}. We obtain
\begin{eqnarray} \label{D1}
\Delta \psi^2(0) &=& \left| \frac{B_s(\nu_p)}{B_s(m)} \right|^2-1
  \,
=
-C_f \als^2 \ln(\als) \bigg\{ \Big[2-\frac23 s(s+1) \Big] C_f
+ C_A \bigg\} 
\\
\nn
&&- \frac{C_f}{\pi} \als^3 \ln^2(\als) \Bigg\{ \frac32 C_f^2 +
  \Big[ \frac{41}{12}-\frac{7}{12} s(s+1) \Big] C_f C_A +\frac23 C_A^2
\\
\nn
&&
\qquad\qquad\qquad\qquad
+\frac{\beta_0}{2} \bigg[ \Big( 2-\frac23 s(s+1) \Big) C_f+C_A \bigg] 
  \Bigg\} + \ldots \,,
\end{eqnarray}
where we have expanded up to second order in
$\ln(\nu_p)=\ln(m\als)$ with $\als\equiv\als(\nu_p)$.  The first
term reproduces the leading log term \cite{BSSCM,leadinglog} 
(see also \cite{HT}), the $\beta_0$-independent
$O(\als^3\ln^2\als)$ terms reproduce Kniehl and Penin results
\cite{KP2} and we agree with the complete $O(\als^3\ln^2\als)$ term
computed by Manohar and Stewart \cite{V1MS} (the sign of difference
for the $\beta_0$-dependent terms displayed in Ref. \cite{V1MS} is due
to the fact that in Ref. \cite{V1MS} the expansion was made with
$\als(m)$ whereas here we have chosen $\als(m\als)$). Nevertheless, 
disagreement with this last evaluation appears at higher orders in the
expansion in $\als$ (we have explicitly checked this for the
$O(\als^4\ln^3\als)$ terms). As far as we can see, the disagreement seems
to be due to the fact that they have different expressions for the RG
improved potentials \cite{vNRQCD2,V1MS}.\footnote{The running of the
Coulomb potential is not needed for the precision of the above
calculation. Nevertheless, the running obtained in vNRQCD
\cite{HMS} also disagrees with the one obtained in pNRQCD \cite{RG}. At
this respect, we would like to report on a recent computation \cite{PS2}
of the 4-loop double log term of the Coulomb potential proportional to
$C_A^3\beta_0$ that agrees with the pNRQCD
result and disagrees with the vNRQCD one.}

By setting $\nu_p \sim m\als$, $B_s(\nu_p)$ includes all the large logs at NLL order in 
any (inclusive enough) S-wave
heavy-quarkonium production observable we can think of. For instance, the decays to 
$e^+e^-$ and to two
photons at NLL order read
\bea
\Gamma(V_Q (nS) \rightarrow e^+e^-) &=&
2 \left[ { \alpha_{em} Q \over M_{V_Q(nS)}} \right]^2
\left({m_QC_f\als \over n}\right)^3
\left\{
B_1(\nu_p)(1+\delta \phi_n)
\right\}^2
\\
\nn
&\simeq&
2 \left[ { \alpha_{em} Q \over M_{V_Q(nS)}} \right]^2
\left({m_QC_f\als \over n}\right)^3
\left\{1+2(B_1(\nu_p)-1)+2\delta \phi_n
\right\}
\,,
\\
\Gamma(P_Q (nS) \rightarrow \gamma\gamma) &=&
6 \left[ { \alpha_{em} Q^2 \over M_{P_Q(nS)}} \right]^2
\left({m_QC_f\als \over n}\right)^3
\left\{
B_0(\nu_p)(1+\delta \phi_n)
\right\}^2
\\
\nn
&\simeq&
6 \left[ { \alpha_{em} Q^2 \over M_{P_Q(nS)}} \right]^2
\left({m_QC_f\als \over n}\right)^3
\left\{1+2(B_0(\nu_p)-1)+2\delta \phi_n
\right\}
\,,
\eea
where $V$ and $P$ stand for the vector and pseudoscalar heavy
quarkonium, we have fixed $\nu_p=m_QC_f\als/n$, $\als=\als(\nu_p)$, and
($\Psi_n(z) ={d^n \ln \Gamma (z) \over dz^n}$ and $\Gamma (z)$ is the
Euler $\Gamma$-function)
\be
\delta \phi_n ={\als \over \pi}
\left[-C_A+
{\beta_0\over 4}
\left(\Psi_1(n+1)-2n\Psi_2(n)+{3 \over 2}+\gamma_E+{2\over n}
\right)
\right]
\,,
\ee
which has been read from Ref. \cite{leadinglog} (see also
\cite{KPPY}). Working along similar lines one could easily obtain NLL
expressions for other heavy quarkonium observables in the study of
$t$-$\bar t$ production near threshold or in sum rules of bottomonium.
Note that for $t$-$\bar t$ production near threshold there already
exists a (partial) NNLL RG improved evaluation within the vNRQCD
formalism \cite{HMST}. Since we disagree for the RG improved expression
for the electromagnetic current matching coefficient, this discrepancy
would also propagate to that evaluation.
   
In conclusion, by using the method of
Ref. \cite{RGmass} and incorporating the idea \cite{LMR} of
correlating the cut-offs of the effective theory, we have given the first steps towards
the creation of a comprehensive system of RG equations in pNRQCD once
the scale $\nu_p$ enters into the game. We have used this
formalism to compute the running of the matching coefficients of the
vector and pseudoscalar currents and disagreement with the results
obtained using the vNRQCD framework \cite{V1MS} has been found. Our
results allow to obtain S-wave heavy quarkonium production observables
with NLL accuracy. We have explicitly illustrated this point for
heavy-quarkonium decays to $e^+e^-$ and to two photons. We have also
computed the $O(m\alpha^8\ln^3\alpha)$ corrections to the Hydrogen
spectrum in the Appendix.

{\bf Acknowledgments}\\
We thank A. Hoang, A.V. Manohar and specially J. Soto for useful
discussions. We also thank J. Soto for comments on the
manuscript.

\section{Appendix: $O(m\al^8\ln^3\al)$ contributions to the Hydrogen energy}

With the above discussion, we may also try to see whether we are able
to obtain the $O(m\al^8\ln^3\al)$ contributions to the Hydrogen
spectrum. It goes beyond the scope of this paper to perform a detailed
analysis. Here, we will just see that under some assumptions from
Ref. \cite{QED}, we are able to obtain the $O(m\al^8\ln^3\al)$
correction to the Hydrogen spectrum.

We will use the notation of Ref. \cite{rgmh}. In that paper,
Hydrogen-like atoms with light fermions were considered. Here, we will
use their results on the strict Hydrogen limit (no light fermions:
$n_f=0$).

According to Ref. \cite{QED}, the $O(m\al^8\ln^3\al)$ correction to
the Hydrogen energy can be obtained from the anomalous dimension due
to diagrams of the type of Fig 2.a in Ref. \cite{QED} or, in our case,
to the diagram in Fig. 1. The argument that led to this conclusion
was that the $O(m\al^8\ln^3\al)$ terms had the highest possible log power
that could appear from a NNNLL evaluation of the energy and that, in order to
achieve such power, it was necessary to mix with NNLL logs. The latter
only appear in the LL evaluation of $D_{d}^{(2)}$, which, indeed, only
produces a single log \cite{QED} (see also \cite{RGmass}). The other point was that the NLL
evaluation of the potentials would only produce single logs unless mixed with LL
running. Therefore, the diagrams with the highest possible power of
$D_{d}^{(2)}$ will give the highest possible log power of the Hydrogen energy at NNNLL.

The RG equation for the coefficient of the delta potential due to Fig. 1 reads
\be
\label{eqDdhyd}
\nu_p {d \over d\nu_p}D_{d}^{(2)}(\nu_p) 
\doteq
 Z^2 \alpha D_{d}^{(2)2}(\nu_p)
\,.
\ee
The loop integral of the diagram of Fig. 1 is just equal to the one
that gives the running of $c_s$ due to $D_{d,s}$. Therefore, we can
obtain Eq. (\ref{eqDdhyd}) from Eq. (\ref{cseq}) by just introducing a
factor 4 due to the fact that we have to change the reduced mass,
$m/2$, from the equal mass case to the reduced mass, $m$, from the
Hydrogen-like case. The left-hand side of Eq. (\ref{eqDdhyd}) gives
the relevant running of $D_{d}^{(2)}$ with NLL accuracy for our
case. Therefore, in the right-hand side we only need $D_{d}^{(2)}$
with LL accuracy, which we read from Ref. \cite{rgmh} in the limit
$n_f=0$:
\be
D_{d}^{(2)}(\nu_p)={\al \over 2}c_D(\nu_p^2/m)\,,
\ee
where
\be
c_D(\nu_p^2/m)=1-{16 \over 3}{\al \over \pi}\ln{\nu_p \over m}
\,.
\ee
Therefore, Eq. (\ref{eqDdhyd}) approximates to
\be
\label{eqDdhydaprox}
\nu_p {d \over d\nu_p}D_{d}^{(2)}(\nu_p) 
\doteq
 Z^2 {\alpha^3 \over 4}c^2_D(\nu_p^2/m)
\,.
\ee
The above equation gives the following correction at $O(\al^5\ln^3)$: 
\be
\delta D_{d}^{(2)} = {64 \over 27}Z^2\al^3 \left({\al \over \pi}\right)^2\ln^3{\nu_p \over m}
\,.
\ee
This contribution gives the following correction to the Hydrogen spectrum 
($\nu_p \sim mZ\al$):
\be
\delta E={64 \over 27}m(Z\al)^6\left({\al \over \pi}\right)^2
{\delta_{l0} \over n^3}\ln^3Z\al
\,.
\ee
This result agrees with the analytical result of Karshenboim
\cite{Karshenboim}, the numerical computation of Goidenko et {\it al.}
\cite{Goidenko} and the analytical result of Manohar and Stewart
\cite{QED}. It disagrees with the numerical computations of Malampalli
and Sapirstein \cite{MalSap} and Yerokhin \cite{Yer1}, which agree
with each other. Nevertheless, it may happen that the latter
computations are not complete for the desired accuracy \cite{Yer2}.

%%%%%%%%%%%%%%%%%%%%% BIBLIOGRAPHY %%%%%%%%%%%%%%%%%%%%%%%%%%%%%%%%%%%%

\end{document}